\documentclass[10pt]{article}
\usepackage{pgfplots}
\pgfplotsset{compat=1.18}
\usepackage[utf8]{inputenc}
\usepackage{newunicodechar,graphicx}
\DeclareRobustCommand{\okina}{%
  \raisebox{\dimexpr\fontcharht\font`A-\height}{%
    \scalebox{0.8}{`}%
  }%
}
\newunicodechar{ʻ}{\okina}
\usepackage{enumitem}
\setlist[itemize,enumerate]{noitemsep, topsep=0pt, leftmargin=1.0em}
\usepackage{background}
\backgroundsetup{
  position=current page.east,
  angle=-90,
  nodeanchor=east,
  vshift=-4mm,
  opacity=0.5,
  scale=3,
  contents=PREPRINT
}
\usepackage[letterpaper]{geometry}
\usepackage{hicss}
\usepackage{times}
\usepackage[none]{hyphenat}
\usepackage{url}
\usepackage{latexsym}
\usepackage{minted}
\usepackage{indentfirst}
\usepackage{graphicx}
\graphicspath{{images/}}
\usepackage[style=apa]{biblatex}
\usepackage{soul}
\usepackage{booktabs}
\usepackage{multirow}
\usepackage{float}
\usepackage{tcolorbox}
\usepackage{tabularx} 
\usepackage{balance}
\usepackage{booktabs}
\usepackage{caption}
\usepackage[hidelinks,bookmarks=true]{hyperref}
\usepackage{colortbl}
\usepackage{fontawesome}
\title{Understanding Computer Science Students' Career Fair Experiences: Goals, Preparation, and Outcomes}

\author{Briana Lee \\
 University of Hawaiʻi at Mānoa \\
 {\underline{\href{mailto:brianall@hawaii.edu}{brianall@hawaii.edu}} } \\ \\
 Kenny Ka‘aiakamanu-Quibilan \\
 University of Hawaiʻi at Mānoa \\
 {\underline{\href{mailto:kennydq@hawaii.edu}{kennydq@hawaii.edu}} } \\ \And
 Samantha Limon \\
 University of Hawaiʻi at Mānoa \\
 {\underline{\href{mailto:splimon@hawaii.edu}{splimon@hawaii.edu}} } \\ \\
Anthony Peruma\\
 University of Hawaiʻi at Mānoa \\
 {\underline{\href{mailto:peruma@hawaii.edu}{peruma@hawaii.edu}} } \\ \And
 Alyssia Chen \\
 University of Hawaiʻi at Mānoa \\
 {\underline{\href{mailto:abc8@hawaii.edu}{abc8@hawaii.edu}} } \\
}

\date{}

\newcommand{\RQA}{\textbf{RQ1}: What are the typical career goals and interests of CS students that attend career fairs?}

\newcommand{\RQB}{\textbf{RQ2}: What are the preparation behaviors and patterns of CS students attending career fairs?}

\newcommand{\RQC}{\textbf{RQ3}: How do CS students evaluate their career fair experience?
}

\newcommand{\RQD}{\textbf{RQ4}: How does the career fair experiences contribute to CS students' career development?
}

\begin{document}
\maketitle
\begin{abstract}
The technology industry offers exciting and diverse career opportunities, ranging from traditional software development to emerging fields such as artificial intelligence, cybersecurity, and data science. Career fairs play a crucial role in helping Computer Science (CS) students understand the various career pathways available to them in the industry. However, limited research exists on how CS students experience and benefit from these events. Through a survey of 86 students, we investigate their motivations for attending, preparation strategies, and learning outcomes, including exposure to new career paths and technologies. We envision our findings providing valuable insights for career services professionals, educators, and industry leaders in improving the career development processes of CS students.

\end{abstract}

\subsubsection*{Keywords:} 

Career Fair, Computer Science Students, Career Preparation, Job Search Strategies, Survey

\section{Introduction}
According to the U.S. Bureau of Labor Statistics, the U.S. has shown significant job growth in the computer and information technology sector. It is projected to continue expanding rapidly, with approximately 356,700 job openings expected each year from 2023 to 2033 due to new positions and workforce turnover (\cite{stats-growth00}). Despite broader economic uncertainties, companies are actively hiring, particularly for specialized positions in fields like artificial intelligence, which continues to see increasing demand (\cite{stats-growth01,stats-growth02}). This growth in the computer and information technology industry presents significant career opportunities for students in Computer Science (CS) and related fields.

However, these opportunities also present unique challenges for CS students. Rapid advancements in technology are creating a demand for skills and roles that are not fully addressed by traditional academic curricula, particularly in areas like cloud computing, artificial intelligence, and data science (\cite{Htte2023,Torosyan2023}). Furthermore, competitiveness in the tech industry can create psychological barriers, such as impostor syndrome, particularly among underrepresented students. (\cite{Chen2024,Rosenstein2020}). 

In this context, university career fairs play a crucial role in connecting students with industry professionals. Through these events, students not only gain insights into potential job openings but also learn about the skills and qualifications employers seek. By participating, students can expand their professional network, enhance their profiles, and increase their chances of securing internships and job offers after graduation in a competitive job market (\cite{Beam2016, Silkes2010}). However, there is little to no research focusing on the experiences and perceptions of CS students as they prepare for and attend career fairs.  

\subsection{Goals and Research Questions}
In this study, we conduct a survey of CS students grounded in Social Cognitive Career Theory (SCCT) (\cite{Lent2006}) with the primary goals of identifying what motivates them to attend career fairs, examining their preparation strategies, and analyzing their experiences and takeaways from these events. SCCT helps understand how outcome expectations, personal goals, and environmental influences affect career development.  We envision our findings not only advancing knowledge in the field of career development for CS students but also providing actionable insights for universities and industry partners. 
We aim to address the following research questions (RQs):
\begin{itemize}
    \item \textbf{\RQA} This RQ aims to establish a baseline understanding of the motivations behind CS students' attendance at career fairs. The findings aim to identify mismatches between student goals and event outcomes.  
    \item \textbf{\RQB} Limited research exists on how CS students prepare for career exploration events. This RQ helps in understanding preparation patterns, best practices, and gaps in student knowledge. The findings can assist in the development of programs to improve student outcomes at these events.
    \item \textbf{\RQC} This RQ aims to assess the student's overall experience at the career fair. Identifying factors that lead to positive evaluations provides actionable insights for improving subsequent career fairs to ensure they are effective for participants. 
    \item \textbf{\RQD} This RQ explores the learning outcomes and  benefits students gain from participating in career fairs. By analyzing the knowledge acquired, advice received, and resources accessed, the findings can provide insights into how to improve career fairs and measure their effectiveness. 
\end{itemize}

\section{Related Work}
In this section, we present a review of related studies.

\textcite{Gordon2014} highlighted that recruits in the hospitality and tourism sectors perceive career fairs positively. Their study found that these events are effective in increasing company visibility and enhancing student preparedness, demonstrating a high level of usefulness. Specifically, the findings indicated a mean rating of 4.39 out of 5, showing that recruits found career fairs to be highly beneficial. They also recommended that students conduct research before attending these events to maximize interactions. Similarly, \textcite{Beam2016} presented a study conducted in the Philippines, which emphasized career fairs as significant opportunities for employment. The survey results indicated that attendance at such fairs had a significant influence on individuals' job-finding abilities. Another important aspect of attending career fairs is the opportunity to develop soft skills and engage in experiential learning. Furthermore, \textcite{Cakir2023} explored the positive outcomes of their university's career fair by assessing employers' perspectives on these events. However, they did not collect data on students' perceptions of career fairs. Their study revealed that 44.7\% of employers had hired graduates they met during career fairs, highlighting the importance of these events in career development and employment. Further, 93.4\% of employers found career fairs beneficial.

\textcite{Buell2020} examined student access to career fairs for online computing students based in Ohio. The authors note that rural institutions are more likely to open their career fairs to the public, providing access for remote students, while urban and online-focused institutions are less inclined to do so. 
In their study, \textcite{Stepanova2021} examined employer recruitment expectations and their outcomes, achieving a 92\% agreement score on the significant relevance of work experience and project-based learning. Additionally, the importance of both technical and soft skills was emphasized, highlighting the industry's need for potential employees to be well-rounded in both technical and collaborative workspaces. Furthermore, \textcite{Chen2025} explored the impact of technological advancements on recruitment processes, particularly the role of generative AI. Their findings indicated that candidates familiar with AI tools such as ChatGPT and GitHub Copilot were viewed as more valuable. Interestingly, while 60.87\% of recruiters reported familiarity with these AI tools, only 15.6\% of organizations had established clear guidelines regarding the use and necessary competencies related to AI. This underscores the need for career fairs and academic curricula to adapt, ensuring that students are prepared to meet evolving industry standards and expectations.
 
\section{Study Design}
\begin{table*}
\centering
\caption[xxx]{Career Fair Survey Questions (\textit{Note: questions ending with a * are mandatory}).}
\vspace{-3mm}
\label{Table:survey}
\resizebox{\textwidth}{!}{%
\begin{tabular}{@{}|l|p{0.45\linewidth}|l|p{0.15\linewidth}|c|@{}}
\toprule
\multicolumn{1}{|c|}{\textbf{No.}} &
  \multicolumn{1}{c|}{\textbf{Survey Question}} &
  \multicolumn{1}{c|}{\textbf{Type}} &
  \multicolumn{1}{c|}{\textbf{Notes}} &
  \textbf{Area} \\ \midrule
\rowcolor[HTML]{ECF4FF} 
1 &
  What is your gender? * &
  Single-Choice &
  Includes ``Other" free-text option &
  \multicolumn{1}{l|}{\cellcolor[HTML]{ECF4FF}} \\ \cmidrule(r){1-4}
\rowcolor[HTML]{ECF4FF} 
2 &
  What is your academic standing? * &
  Single-Choice &
   & 
  \multicolumn{1}{l|}{\cellcolor[HTML]{ECF4FF}} \\ \cmidrule(r){1-4}
\rowcolor[HTML]{ECF4FF} 
3 &
  What is/was your primary field of study/major? * &
  Single-Choice & 
  Includes ``Other" free-text option & 
   \multirow{-3}{*}{\cellcolor[HTML]{ECF4FF}\rotatebox[origin=c]{90}{Demographic}} \\ \midrule
\rowcolor[HTML]{FEE7E6} 
4 &
  What was your primary goal for attending this career fair? * &
  Single-Choice &
  Includes ``Other" free-text option &
  \cellcolor[HTML]{FEE7E6} \\ \cmidrule(r){1-4}
\rowcolor[HTML]{FEE7E6} 
5 &
  Choose the top 3 types of career opportunities based on your level of interest * &
  Multi-Choice &
 Includes ``Other" free-text option &
  \multirow{-2}{*}{\cellcolor[HTML]{FEE7E6}\rotatebox[origin=c]{90}{RQ 1}} \\ \midrule
\rowcolor[HTML]{ECF4FF} 
6 &
  To what extent did you prepare for this career fair? * &
  Single-Choice &
   &
  \cellcolor[HTML]{ECF4FF} \\ \cmidrule(r){1-4}
\rowcolor[HTML]{ECF4FF} 
7 &
  How much time did you  spend preparing for the career fair? * &
  Single-Choice &
   &
  \cellcolor[HTML]{ECF4FF} \\ \cmidrule(r){1-4}
\rowcolor[HTML]{ECF4FF} 
8 &
  How did you prepare for this career fair? * &
  Multi-Choice &
  Includes ``Other" free-text option &
  \cellcolor[HTML]{ECF4FF} \\ \cmidrule(r){1-4}
\rowcolor[HTML]{ECF4FF} 
9 &
  How would you rate  your level of preparation for the career fair? * &
  Single-Choice &
   &
  \multirow{-4}{*}{\cellcolor[HTML]{ECF4FF}\rotatebox[origin=c]{90}{RQ 2}} \\ \midrule
\rowcolor[HTML]{FEE7E6} 
10 &
  Approximately how many companies did you interact with at the career fair? * &
  Number &
   &
  \cellcolor[HTML]{FEE7E6} \\ \cmidrule(r){1-4}
\rowcolor[HTML]{FEE7E6} 
11 &
  How confident are you that you will receive a follow-up from one or more companies you interacted  with at the career fair? * &
  Single-Choice &
   &
  \cellcolor[HTML]{FEE7E6} \\ \cmidrule(r){1-4}
\rowcolor[HTML]{FEE7E6} 
12 &
  Rate your overall   experience at the career fair * &
  Single-Choice &
   &
  \cellcolor[HTML]{FEE7E6} \\ \cmidrule(r){1-4}
\rowcolor[HTML]{FEE7E6} 
13 &
  What were the main factors that influenced your rating of the career fair? &
  Free Text &
   &
  \multirow{-4}{*}{\cellcolor[HTML]{FEE7E6}\rotatebox[origin=c]{90}{RQ 3}} \\ \midrule
\rowcolor[HTML]{ECF4FF} 
14 &
  Did you learn about new career paths or technologies that you weren't aware of before? * &
  Single-Choice &
  ``Yes'' / ``No'' answer options & 
  \cellcolor[HTML]{ECF4FF} \\ \cmidrule(r){1-4}
\rowcolor[HTML]{ECF4FF} 
15 &
  Please provide some examples of new career paths or technologies you weren't aware of before * &
  Free Text &
  Shown if ``Yes'' is chosen above &
  \cellcolor[HTML]{ECF4FF} \\ \cmidrule(r){1-4}
\rowcolor[HTML]{ECF4FF} 
16 &
  What was the most valuable piece of information or advice you received at the career fair? &
  Free Text &
   &
  \cellcolor[HTML]{ECF4FF} \\ \cmidrule(r){1-4}
\rowcolor[HTML]{ECF4FF} 
17 &
  What additional resources or support would have enhanced your career fair experience? &
  Free Text &
   &
  \multirow{-4}{*}{\cellcolor[HTML]{ECF4FF}\rotatebox[origin=c]{90}{RQ 4}} \\ \bottomrule
\end{tabular}%
}
\end{table*}

The research strategy for this study employs a respondent-driven design, utilizing a sample survey methodology (\cite{Stol2018}). This approach involves gathering data from a broad and diverse group of computing and information technology students through an online questionnaire. Below, we describe the key activities in our methodology. Artifacts for this study are available at: \url{https://doi.org/10.6084/m9.figshare.29327414}

\subsection{Survey Design}
Our survey included 17 questions that focused on participants' demographics, their preparation for the career fair, and their experiences. Following best practices (\cite{Kitchenham2002,kasunic2005designing}), the questions were based on Social Cognitive Career Theory and aligned with the objectives of our study. We utilized Qualtrics to design and distribute the survey, allowing only one response per participant. Additionally, we conducted a pilot run of the survey with five participants, which helped us identify questions that needed to be rephrased, reorganized, and removed/replaced to improve understandability\footnote{Results from the pilot run were discarded and did not contribute to the final analysis.}. Table \ref{Table:survey} shows the questions included in our survey. The complete questionnaire, as presented to participants, is available at: \url{https://doi.org/10.6084/m9.figshare.29327414}.

\subsection{Sampling Strategy and Recruitment}
We utilized a convenience sampling approach (\cite{Baltes2022}) to recruit participants for our survey. Specifically, we surveyed students who attended the Fall 2024 career fair hosted by the Department of Information Computer Sciences at the University of Hawaiʻi at Mānoa. This career fair is held twice a year and is specifically designed for students seeking tech internships and full-time positions. The event featured 29 organizations from a diverse range of sectors, including government and public service, health and human services, finance and insurance, education, tech, and defense. 

We approached students as they left the venue and provided them with handouts containing a QR code for the online survey. The survey was open for only one week after the event, allowing us to gather feedback while their experiences were still fresh in their minds. Most completed it immediately on their smartphones, while others did so later. This method aimed to improve the quality and relevance of the data collected and minimize recall bias that could arise if participants completed the survey weeks after the event.

\subsection{Data Analysis}
We utilized a mixed-methods strategy for data analysis, involving both quantitative and qualitative approaches (\cite{wagner2020challenges}). The quantitative analysis used statistical methods, while a thematic analysis was conducted on open-ended responses (\cite{Braun2006}). Three authors independently analyzed and coded the responses, identifying emerging themes and resolving any differences through discussion. In some cases, responses included multiple themes. Further, some responses were excluded from the thematic analysis due to ambiguity or irrelevance.

\section{Results}
In this section, we address our RQs based on the survey results. We received a total of 120 responses to the survey; however, not all participants answered all mandatory questions. To ensure consistency, our analysis focuses only on those participants who responded to all required questions, resulting in 86 valid responses. Therefore, the following results are based only on these 86 responses.

\subsection*{\textbf{Participant Demographics}}
Before addressing the RQs, we provide an overview of the demographics of our survey participants. Out of the total 86 participants, 57 individuals (66.28\%) identified as male, 27 as female, and two as non-binary or third gender. Regarding academic standing, the majority were seniors, totaling 40 participants (46.51\%), followed by five freshmen, 16 sophomores, and 15 juniors. Further, seven participants were identified as graduate students and three as alumni. Finally, most participants majored in Computer Science (70 participants), followed by nine in Management Information Systems and four in Engineering.

Although our participants are limited to a single institution, these demographic details indicate that they are well-suited for the study. The high concentration of CS students indicates relevant expertise in the subject matter, which improves the validity of our findings.

\subsection{RQ1: What are the typical career goals and interests of CS students that attend career fairs?}

To address RQ1, we analyzed responses to questions \#4 and \#5, which explored participants' primary goals for attending the career fair and their top three career opportunities of interest.

In response to question \#4, which asked participants to identify their primary goal for attending the career fair, 45 participants (52.33\%) indicated that they were looking for internship opportunities, while 25 individuals (29.07\%) sought full-time job opportunities. Four participants (4.65\%) expressed a desire to explore different career paths, six (6.98\%) wanted to learn more about various companies, and another six (6.98\%) aimed to network with industry professionals.

These results indicate that an overwhelming majority of participants were primarily focused on seeking internship opportunities, with more than half of the participants citing this as their main goal for attending the fair. The high percentage of participants looking for internships suggests that many attendees were early in their academic careers and eager to gain work experience. When we break down the data by class standing, those seeking internships included 4 freshmen, 14 sophomores, 12 juniors, 14 seniors, and 1 graduate student. This indicates that nearly all freshmen (4 out of 5), sophomores (14 out of 16), and juniors (12 out of 15) who attended the fair were primarily motivated by the prospect of internships. In contrast, participants pursuing full-time roles were predominantly seniors (18 individuals), along with 2 alumni and 5 graduate students. These findings are intuitive, as seniors and graduate students are typically more focused on preparing to commence full-time employment. Overall, the results reveal a clear relationship between class standing and primary goals for attending the career fair: underclassmen are more internship-oriented, while upperclassmen and graduates tend to focus more on securing full-time employment.

In question \#5, participants were asked to select three types of career opportunities they were most interested in. As shown in Table \ref{tab:career-interest}, Software Development was the most popular career preference, chosen by 55 participants (64.0\% of participants). This was followed closely by Data Science with 47 selections (54.7\%), cybersecurity with 36 selections (41.9\%), and Artificial Intelligence/ML (AI/ML) with 30 selections (34.9\%). Additionally, software development and data science are often selected together, appearing in 29 responses. Additionally, AI/ML is frequently paired with either Data Science or Software Development.  A concerning observation is the low interest in Quality Assurance/Testing (QA), with only six selections. This lack of interest is troubling given that QA is essential to software development, highlighting a potential knowledge gap regarding the roles of QA professionals.

While non-coding roles appear less frequently than coding roles, we observe some interesting patterns, such as Business Analyst being often paired with Data Science (13 times), suggesting that participants view Business Analysts as essential links between data insights and business decisions. We also observed eight instances where Business Analysts and Project Management roles were combined, demonstrating how these roles naturally complement each other.

Finally, among the five participants who responded with ``Other,'' three stated they were interested in game development, while one each expressed interest in data engineering and hardware engineering.

\begin{table}
\centering
\caption[xxx]{Distribution of career preferences.}
\vspace{-3mm}
\label{tab:career-interest}
\resizebox{\columnwidth}{!}{%
\begin{tabular}{@{}p{0.66\linewidth}rr@{}}
\toprule
\multicolumn{1}{c}{\textbf{Career Field}} & \multicolumn{1}{c}{\textbf{Count of Selections}} \\ \midrule
Software Development       & 55 \\ \midrule
Data Science               & 47 \\ \midrule
Cybersecurity              & 36 \\ \midrule
Artificial Intelligence/ML & 30 \\ \midrule
Business Analyst           & 19 \\ \midrule
Network Engineering        & 18 \\ \midrule
UX/UI Design               & 16 \\ \midrule
Project Management         & 16 \\ \midrule
DevOps                     & 9  \\ \midrule
Quality Assurance/Testing  & 6  \\ \midrule
Other                      & 5  \\ \midrule
Help Desk Technician       & 1  \\ \bottomrule
\end{tabular}%
}
\end{table}

These results indicate that software engineering remains the primary career aspiration among computer science students attending career fairs, aligning with broader industry trends. However, the growing interest in fields such as cybersecurity and AI/ML suggests that students are increasingly drawn to newer areas within the technology sector. Additionally, the inclusion of business analysis among students' preferences highlights an interest in roles that bridge technology and business. This indicates a trend toward interdisciplinary career paths in the tech industry, which require analytical skills that extend beyond purely technical capabilities.

\noindent\faInfoCircle\hspace{0.3em}\textbf{\textit{Summary for RQ1.}}
The results indicate that most underclassmen studying Computer Science attend career fairs primarily to secure internships. In contrast, seniors and graduate students tend to focus more on finding full-time positions. When it comes to career interests, software engineering is the most sought-after field, followed by a growing interest in cybersecurity, AI/ML, and business analysis. This trend suggests that students are not only pursuing traditional tech roles but are also increasingly exploring emerging career paths.

\subsection{RQ2: What are the preparation behaviors and patterns of CS students attending career fairs?}

To address this RQ, we analyzed responses to questions \#6 through \#9, which examined how participants prepare for the career fair.

For question \#6, nearly half of the 40 participants (45.51\%) indicated that they prepared a ``Moderate amount,'' while 28 participants (32.56\%) reported doing ``A little'' preparation. Five participants indicated they prepared ``A lot,'' and only 2 participants (2.33\%) said they prepared ``A great deal.” Meanwhile, 11 participants (12.79\%) stated they did ``None at all.'' This suggests that most participants engage in little to moderate preparation for the career fair, with only a slight majority making intensive preparations. 

Question \#7 examines the time spent preparing for the career fair.  36 participants (41.86\%) devoted ``1-3 hours'' to preparation, while 27 participants (31.40\%) spent ``Less than 1 hour.'' Further, 15 (17.44\%) reported ``3-5 hours'' of preparation. 

When we analyzed the data by class standing, we found that four out of five Freshmen indicated they had prepared ``A little'' to ``A moderate amount'' for the career fair. Additionally, nine out of sixteen Freshmen spent ``A little'' time preparing, which they reported as ``Less than an hour'' or ``1-3 hours.'' Eight of 15 juniors and 19 of 40 seniors spent ``A moderate amount'' of time preparing, which equated to ``1-3 hours.'' In contrast, seven graduate students spend a ``Moderate amount'' of time preparing, which they reported as ``Less than 1 hour'' to ``3-5 hours.''

In regard to the preparation activities outlined in question \#8, participants typically selected an average of two preparation methods, as shown in Table \ref{tab:prep-methods}. The most common approach, chosen by 68 participants (46.9\%), was ``Created/Updated resume.'' The second most common method was ``Researched companies in advance,'' with 36 responses (24.8\%). The third most frequent activity was ``Attended career workshops,'' with 16 responses (11.0\%). Interestingly, there were no responses for ``Asked for advice from faculty, the career office, or other university representatives.'' The 4 ``Other'' responses noted that they did not prepare or had just found out about the career fair.

\begin{table}
\centering
\caption[xxx]{Distribution of career fair preparation methods.}
\vspace{-3mm}
\label{tab:prep-methods}
\resizebox{\columnwidth}{!}{%
\begin{tabular}{@{}p{0.66\linewidth}rr@{}}
\toprule
\multicolumn{1}{c}{\multirow{2}{*}{\textbf{Preperation Method}}}                 & \multicolumn{2}{c}{\textbf{Responses}}                                       \\ \cmidrule(l){2-3} 
\multicolumn{1}{c}{}                                                             & \multicolumn{1}{c}{\textbf{Count}} & \multicolumn{1}{c}{\textbf{Percentage}} \\ \midrule
Created/Updated resume          & 68 & 46.90\% \\ \midrule
Researched companies in advance & 36 & 24.83\% \\ \midrule
Attended career workshops       & 16 & 11.03\% \\ \midrule
Practiced interview questions   & 11 & 7.59\%  \\ \midrule
Prepared elevator pitch         & 10 & 6.90\%  \\ \midrule
Other                           & 4  & 2.76\%  \\ \midrule
Asked for advice from faculty, career office, or other university representative & 0                                  & 0.00\%                                  \\ \bottomrule
\end{tabular}%
}
\end{table}

In question \#9, we asked participants to evaluate their readiness for the career fair. The results showed that 30 participants (34.88\%) rated their preparedness as ``Good,'' making this the most common answer. Additionally, 19 participants (22.09\%) rated their readiness as ``Very Good,'' while three participants (3.49\%) chose ``Excellent.'' Conversely, 23 participants (26.74\%) rated their preparedness as ``Fair,'' and 11 (12.79\%) indicated they felt ``Poorly'' prepared.

\noindent\faInfoCircle\hspace{0.3em}\textbf{\textit{Summary for RQ2.}}
The results indicate that students were generally prepared lightly to moderately for the tech career fair. They usually spent one to two hours preparing, with their main activities including updating their resumes and researching companies. When asked to rate their readiness, the majority of students placed themselves between ``Fair'' and ``Good,'' while a few felt exceptionally well-prepared or unprepared, suggesting that most participants felt adequately ready for the event.

\subsection{RQ3: How do CS students evaluate their career fair experience?}

To address this RQ, we analyzed responses to survey questions \#10 through \#13. These questions focused on the number of companies participants interacted with, their confidence in receiving follow-up communications, their overall rating of the event, and the factors that influenced their ratings.

\begin{figure}[b]
\begin{center}
\caption{Number of companies visited per participant.}
\label{career-fair-interactions}
\vspace{-3mm}
\begin{tikzpicture}
\begin{axis}[
    ybar,
    xlabel={Number of Companies},
    ylabel={Participants},
    xtick=data,
    enlarge x limits=0.05,
    symbolic x coords={1,2,3,4,5,6,7,8,9,10,15},
    ymin=0,
    bar width=10pt,
    nodes near coords,
    width=0.5\textwidth,
    height=6cm
]
\addplot coordinates {
    (1,4) (2,7) (3,7) (4,11) (5,24) (6,8) (7,8) (8,6) (9,2) (10,7) (15,2)
};
\end{axis}
\end{tikzpicture}
\end{center}
\end{figure}
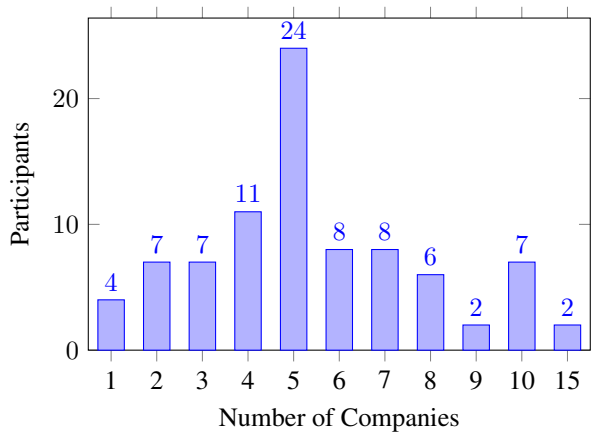

Question \#10 asked participants to report how many companies they interacted with during the career fair. As shown in Figure \ref{career-fair-interactions}, 
the most common response was five companies, reported by 24 participants (27.91\%), followed by 11 participants who visited four companies. The median number of companies visited by participants was five, which represents approximately 17.24\% of the total companies that participated in the event. While this figure might initially seem low, it is important to consider various factors that may have influenced student engagement, such as time constraints, crowd size, and a targeted approach toward companies that aligned with their career goals.

Survey question \#11, which utilized a Likert scale, assessed participants' confidence in receiving follow-up communication from one or more of the companies they interacted with. The majority of participants reported low to moderate confidence levels. Specifically, 16 participants (18.60\%) indicated they were ``Not at all confident,'' while 26 participants (30.24\%) selected ``Slightly confident.'' A total of 30 participants (34.88\%) considered themselves ``Moderately confident.'' Fewer participants expressed higher levels of confidence, with 9 participants (10.47\%) reporting they were ``Very confident'' and 5 participants (5.81\%) stating they were ``Extremely confident.'' These results suggest that although most participants engaged with a diverse range of employers, many participants left the event feeling uncertain about the outcome.

In survey question \#12, participants evaluated their overall experience at the career fair using a five-point Likert scale. Only one participant (1.16\%) rated their experience as ``Poor,'' while 10 participants (11.63\%) selected ``Fair.'' A larger number rated the event as ``Good'' (36 participants, 41.86\%), ``Very Good'' (26 participants, 30.23\%), or ``Excellent'' (13 participants, 15.12\%). These ratings reflect a generally positive perception of the event, with the majority of participants rating their experience as ``Good'' or higher.

To understand the ratings, we analyzed the open-ended responses to question \#13, which asked participants to describe the main factors influencing their experience. Out of 86 participants, 36 provided feedback. Their responses revealed both positive and critical themes. Below are the positive themes identified:

\begin{itemize}
    \item \textbf{Diverse Opportunities:} Participants valued the variety and diversity of employers present at the fair, as well as the range of opportunities available. One participant mentioned, ``The ample space and informational flyers were helpful in learning about what opportunities were available and where each company was.''  Those who highlighted this aspect generally rated the fair as ``Good'' or higher.
    \item \textbf{Positive Interactions:} Many participants expressed their appreciation for the welcoming and friendly nature of the employers, which enhanced their overall experience at the fair. For instance, one participant noted,  ``All the company representatives were nice and approachable!'' These positive interactions built feelings of encouragement, particularly for students attending their first career fair. The ratings from these participants ranged from ``Fair'' to ``Excellent.'' 
    \item \textbf{Informational Resources:} Five participants highlighted the value of the materials and information available at the fair. They found the job postings, event layouts, and flyers to be helpful in navigating and discovering options they were previously unaware of. As one participant noted, ``...flyer given was helpful in learning about what opportunities were available and where each company was.''
    \item \textbf{Networking:} Participants valued their opportunity to engage in direct conversations with professionals, allowing them to establish early connections in the tech industry. Five participants highlighted networking as a key aspect of their experience. As one participant mentioned, ``Had a good experience discussing opportunities with the employers and getting to network with them.''
    \item \textbf{Logistics:} The venue and atmosphere of the fair influenced student perceptions, as a participant noted, “The people at each booth and the size of the venue were both adequate,” and rated the fair as ``Good.'' However, precautions should be taken regarding scheduling to prevent fatigue when multiple fairs are organized in close succession. As another participant mentioned, ``Since it came after the other career fair, it was less hyped.''    
    \item \textbf{Peer Support:} Participants also noted the presence of friends as an important factor. Attending with peers made them feel more at ease and confident, as one expressed, ``Friends were there. Friendly faces.''
\end{itemize}

\noindent Below are the negative themes identified for \#13:

\begin{itemize}
    \item \textbf{Limited Employer Variety:} Participants expressed concern over the limited number of employers at the fair, noting that the same companies tend to participate in every event. As one participant remarked,  ``It’s the same companies every year; there’s little purpose in returning.''    
    \item \textbf{Mismatch of Interests:} Two participants noted that the companies available did not align with their career goals or skill sets. As one participant mentioned, ``Many companies were looking for roles that I was not suited for,'' and rated their experience as ``Fair.'' 
    \item \textbf{Limited Preparation Time:} A participant mentioned having limited preparation time, stating they ``Didn’t have enough time to prepare due to busy schedule.''
\end{itemize}

\noindent\faInfoCircle\hspace{0.3em}\textbf{\textit{Summary for RQ3.}}
CS students generally had positive experiences at the career fair, with most interacting with around five companies and rating the event as ``Good'' or higher. Key factors contributing to satisfaction included diverse opportunities, positive interactions with employers, helpful informational resources, networking opportunities, and adequate logistics. Peer support also made some attendees feel more comfortable. Areas for improvement include limited employer variety, mismatches between company offerings and student interests, and insufficient preparation time for some participants.

\subsection{RQ4: How does the career fair experience contribute to CS students' career development?}
We address this RQ by examining the responses to survey questions \#14 through \#17.

In analyzing the Yes/No responses to question \#14, approximately half of the participants, 46 (53.49\%), indicated that they learned about new career paths or technologies by attending the career fair. In contrast, 40 participants (46.51\%) reported that they did not learn anything new. Those who answered affirmatively were then asked question \#15, which asked them to share examples of the new career paths or technologies they discovered at the career fair. Below are the identified themes from the responses to question \#15:

\begin{itemize}
    \item \textbf{Technical Careers:} In this theme, 21 participants discovered roles in software development, product management, and specialized fields like security and data analysis. One participant noted, ``I wasn't aware that there were so many variations of IT.''
    \item \textbf{Government / Defense:} Several participants found Department of Defense positions to be more diverse than they had previously thought. One participant noted that there were ``different types of paths in the FBI (intelligence analysts, data analysts, etc).''
    \item \textbf{Management \& Consulting:} Thirteen participants highlighted the application of technical skills in roles like consulting, project management, and business analysis. One participant noted, ``I could work as a product manager for Google, which is certainly a career path I never considered.''
\end{itemize}

Survey question \#16 asked participants about the most valuable information or advice they received at the career fair. Below are the identified themes:
\begin{itemize}
    \item \textbf{Job Search and Application Strategies:} This theme was mentioned by 14 participants, who emphasized the importance of being proactive in job applications and networking. For instance, one participant noted, ``Be sure to start working on things once the opportunity to apply for positions opens up.''
    \item \textbf{Resume Customization:} Another important theme emphasized by nine participants was the need to customize resumes for specific job opportunities. As one participant stated, ``Tailor your resume to specific jobs/internships.'' 
    \item \textbf{Company Details and Hiring Expectations:} 
   This theme involves networking with employers to gain insights about the specific company and its hiring expectations. As one participant noted, ``learned about what opportunities were coming up and where to find them in the future.''
    \item \textbf{Soft Skill Development:} Participant highlighted the importance of soft skills and acquiring new technical abilities, stating,  ``Take risks and put yourself out there. It's better to try than regret that you didn't.''  
\end{itemize}

Finally, question \#17 explored additional resources or support that would have enhanced the career fair experience. Identified themes include:
\begin{itemize}
    \item \textbf{Expanding Employer and Company Representation:} Participants expressed a need for greater representation from a variety of businesses, including requests for ``more DoD and federal employers'' and ``big tech'' companies. Additionally, some participants asked for more detailed information about companies and clearer expectations from employers.
    \item \textbf{Resume Preparation:} This theme highlights the importance of support in preparing resumes and emphasizes the value of employer feedback to enhance students' experiences at career fairs. For instance, one participant mentioned, ``...I wish more employers would have accepted my resumes so I could receive feedback on where I stand.''
   \item \textbf{General Preparation:} This theme emphasizes the need to educate participants about career fairs and their purpose. It includes effective preparation strategies, such as developing an elevator pitch. As one participant noted, it’s essential to understand ``how to maximize the effectiveness of first meetings...''.   
\end{itemize}

\noindent\faInfoCircle\hspace{0.3em}\textbf{\textit{Summary for RQ4.}}
Career fairs help CS students by exposing them to new career paths, technologies, and advice. Students discovered roles in areas like cybersecurity and product management, gained job search strategies, and learned the importance of resume customization and soft skills. However, they suggested improvements, such as broader employer representation and on-site resume feedback to improve the effectiveness of career fairs.

\section{Discussion}
This study provides an in-depth examination of how CS students participate in a tech-focused career fair. 
Our findings show that these events support CS students' professional development by providing immediate and long-term opportunities. Many students aimed to secure internships or jobs, while others sought insights into the tech industry, highlighting these events as recruitment drives and learning experiences. For some, it was their first exposure to career fields they had not previously been aware of, which reinforces the importance of these events as spaces for career discovery.

However, preparation strategies varied widely. While some students invested significant time refining their resumes and researching companies, others attended the fair with little to no preparation. This disparity affected confidence levels and satisfaction. These findings suggest that universities should provide more structured guidance on how to approach career fairs, particularly for first-time attendees.

Employer representation also played a significant role in supporting CS students’ career development by offering both exposure and guidance. Over half of the participants discovered new career paths, such as IT support, cybersecurity, and product management. Additionally, many participants appreciated the chance to meet face-to-face with employers, the positive interactions they experienced, the networking opportunities, the resources provided, and the overall organization of the event. These factors contributed to participants' satisfaction and their perception of the fair, highlighting the importance of featuring a diverse range of companies and effectively communicating with students about their expectations. However, while many appreciated the diversity of industries represented, some attendees, especially those who had participated in previous fairs, expressed disappointment over the repeated presence of the same employers at every event.

\subsection{Implications}
\noindent\textbf{For Academic Institutions:} The findings emphasize the need for structured career preparation programs in academic institutions tailored to students' academic levels. These should include resume workshops, mock interviews, and guidance on employer research. Additionally, workshops on soft skills would help students to interact confidently with employers in professional settings. Institutions should also support underrepresented students through mentorship programs or affinity group workshops to promote inclusivity and encourage participation in career fairs. Furthermore, organizers should carefully consider the academic calendar when scheduling career fairs. Hosting a targeted CS fair too soon after a larger all-campus event may reduce its visibility and impact. 

Moreover, with the rapid advancements in the technology sector, it is essential for institutions to collaborate with industry partners to better understand employer expectations, thereby better preparing students for the workforce. This collaboration can be achieved by forming advisory boards made up of industry professionals who can provide insights into the skills that employers value most, leading to improved course curricula. Additionally, integrating guest lectures, internships, and networking opportunities into these programs can further improve students’ real-world exposure and professional connections.

\noindent\textbf{For Employers:} Employers can maximize their impact at career fairs by clearly communicating job roles and expectations for the hiring process. Offering on-site resume feedback and brief skill demonstrations can also be beneficial. Furthermore, employers could collaborate with the university to ensure that the job opportunities available at their company align with students' interests. This approach will help increase the likelihood of successfully recruiting qualified candidates.

\section{Threats to Validity}
With all participants affiliated with a single institution, our results may not be directly generalizable to a broader population. Additionally, to minimize subjectivity in our analysis of free-text responses, three authors independently categorized the reactions and then reached a consensus. There is also a possibility that participants interpreted the survey questions differently, which may have led to variations in their responses. Furthermore, since our study did not include follow-up interviews with survey participants, we lack detailed explanations for some of their survey answers, which could have provided more profound insights into their experiences. However, the anonymous nature of our survey encouraged participants to provide honest and unbiased responses. Lastly, our research scope may not encompass all aspects of how students prepare for and participate in career fairs, which poses a threat to the internal validity of our study. Even with these challenges, our work serves as an exploratory study, aiming to identify preliminary trends and patterns that can inform future research in this area.

\section{Conclusion \& Future Work}
Career fairs play a crucial role in supporting CS students' career development by offering opportunities for internships, full-time positions, and exposure to new career paths. This study highlights the importance of these events as both recruitment platforms and learning opportunities by surveying 86 career fair participants. While many participants appreciated the opportunity to network, gain insights into the tech industry, and receive valuable advice, the level of preparation varied significantly, affecting their confidence and satisfaction. 

Since our study is limited to a single institution, our future work will expand to incorporate multiple institutions across diverse geographic and institutional contexts. This broader approach will enable stronger comparisons and enhance generalizability. Additionally, we aim to track longitudinal outcomes such as internship conversions and job offers, to gain deeper insights into the long-term impacts of career fair participation.

\printbibliography

\end{document}